\documentclass[twocolumn,superscriptaddress,pra,aps]{revtex4}
\usepackage{amsmath,mathrsfs,amsbsy,color,graphicx,bm,amsthm,amsfonts,dsfont}
\usepackage{times}
\usepackage{braket}
\usepackage{multirow}
\usepackage{units}

\begin{document}
\title{Experimental quantum channel simulation}

\author{He Lu}
\affiliation{Shanghai Branch, National Laboratory for Physical Sciences at Microscale and Department of Modern Physics, University of Science and Technology of China, Shanghai 201315, China}
\affiliation{Synergetic Innovation Center of Quantum Information and Quantum Physics, University of Science and Technology of China, Hefei, Anhui 230026, China}
\author{Chang Liu}
\affiliation{Shanghai Branch, National Laboratory for Physical Sciences at Microscale and Department of Modern Physics, University of Science and Technology of China, Shanghai 201315, China}
\affiliation{Synergetic Innovation Center of Quantum Information and Quantum Physics, University of Science and Technology of China, Hefei, Anhui 230026, China}
\author{Dong-Sheng Wang}
\affiliation{Institute for Quantum Science and Technology, University of Calgary, Alberta T2N 1N4, Canada}
\author{Luo-Kan Chen}
\affiliation{Shanghai Branch, National Laboratory for Physical Sciences at Microscale and Department of Modern Physics, University of Science and Technology of China, Shanghai 201315, China}
\affiliation{Synergetic Innovation Center of Quantum Information and Quantum Physics, University of Science and Technology of China, Hefei, Anhui 230026, China}
\author{Zheng-Da Li}
\affiliation{Shanghai Branch, National Laboratory for Physical Sciences at Microscale and Department of Modern Physics, University of Science and Technology of China, Shanghai 201315, China}
\affiliation{Synergetic Innovation Center of Quantum Information and Quantum Physics, University of Science and Technology of China, Hefei, Anhui 230026, China}
\author{Xing-Can Yao}
\affiliation{Physikalisches Institut, Ruprecht-Karls-Universit\"{a}t Heidelberg, Im Neuenheimer Feld 226, 69120 Heidelberg, Germany}
\author{Li Li}
\affiliation{Shanghai Branch, National Laboratory for Physical Sciences at Microscale and Department of Modern Physics, University of Science and Technology of China, Shanghai 201315, China}
\affiliation{Synergetic Innovation Center of Quantum Information and Quantum Physics, University of Science and Technology of China, Hefei, Anhui 230026, China}
\author{Nai-Le Liu}
\affiliation{Shanghai Branch, National Laboratory for Physical Sciences at Microscale and Department of Modern Physics, University of Science and Technology of China, Shanghai 201315, China}
\affiliation{Synergetic Innovation Center of Quantum Information and Quantum Physics, University of Science and Technology of China, Hefei, Anhui 230026, China}
\author{Cheng-Zhi Peng}
\affiliation{Shanghai Branch, National Laboratory for Physical Sciences at Microscale and Department of Modern Physics, University of Science and Technology of China, Shanghai 201315, China}
\affiliation{Synergetic Innovation Center of Quantum Information and Quantum Physics, University of Science and Technology of China, Hefei, Anhui 230026, China}
\author{Barry C. Sanders}
\affiliation{Shanghai Branch, National Laboratory for Physical Sciences at Microscale and Department of Modern Physics, University of Science and Technology of China, Shanghai 201315, China}
\affiliation{Synergetic Innovation Center of Quantum Information and Quantum Physics, University of Science and Technology of China, Hefei, Anhui 230026, China}
\affiliation{Institute for Quantum Science and Technology, University of Calgary, Alberta T2N 1N4, Canada}
\affiliation{Program in Quantum Information Science, Canadian Institute for Advanced Research, Toronto, Ontario M5G 1Z8, Canada}
\author{Yu-Ao Chen}
\affiliation{Shanghai Branch, National Laboratory for Physical Sciences at Microscale and Department of Modern Physics, University of Science and Technology of China, Shanghai 201315, China}
\affiliation{Synergetic Innovation Center of Quantum Information and Quantum Physics, University of Science and Technology of China, Hefei, Anhui 230026, China}
\author{Jian-Wei Pan}
\affiliation{Shanghai Branch, National Laboratory for Physical Sciences at Microscale and Department of Modern Physics, University of Science and Technology of China, Shanghai 201315, China}
\affiliation{Synergetic Innovation Center of Quantum Information and Quantum Physics, University of Science and Technology of China, Hefei, Anhui 230026, China}

\begin{abstract}
Quantum simulation is of great importance in quantum information science. 
Here, we report an experimental quantum channel simulator imbued with an algorithm
for imitating the behavior of a general class of quantum systems. 
The reported quantum channel simulator consists of four single-qubit gates and one controlled-NOT gate.
All types of quantum channels can be decomposed by the algorithm and implemented on this device.
We deploy our system to simulate various quantum channels,
such as quantum-noise channels and weak quantum measurement.
Our results advance experimental quantum channel simulation, 
which is integral to the goal of quantum information processing. 
\end{abstract}

\maketitle

\section{Introduction}
Quantum simulation~\cite{Fey82,Llo96,BN09}
is the most promising near-term application of quantum computing
due to the resource requirements for imitating some classically intractable systems being
significantly less onerous that for other applications such as factorization.
Experimental quantum simulation on closed systems is well studied using photons~\cite{Aspuru2012},
atoms~\cite{CZ12} and trapped ions~\cite{Blatt2012}.
Quantum simulation of open-system dynamics
also has variety of applications, such as
dissipative quantum phase transitions~\cite{PI11}
and dissipative quantum-state engineering~\cite{Diehl08},
thermalization~\cite{TD00},
quantum noise generators~\cite{Lu2014},
non-Markovian dynamics~\cite{Liubiheng2011},
and non-unitary quantum computing~\cite{VWC09}.
Although any non-unitary quantum dynamics
could be embedded into unitary dynamics over a larger Hilbert space
with Hamiltonian evolution~\cite{NC00},
such a direct approach entails a large
computational-space overhead and resultant experimental complexity.
Previous experiments have demonstrated 
a universal unitary-gate on a reprogrammable waveguide chip~\cite{Carolan2015}
and have explored some open-system single-qubit dynamics
such as quantum noise~\cite{Jeong13,Shaham12,shaham11},
weak measurement~\cite{Kim2011,Kim09}
and transpose~\cite{Sciarrino04,Lim11pra, Lim11}.

Our approach is quite distinct from these prior achievements in that our one apparatus
simulates all these transformations and more,
in fact \emph{any} single-qubit channel. We aim to realize a digital single-qubit
channel simulator,
which will serve as reconfigurable component of a nonunitary quantum circuit
that would simulate nonunitary circuits.
Furthermore our quantum simulator is ``digital''. 
A digital quantum simulator is more versatile, as it is able to simulate a wider range of Hamiltonians\cite{CZ12}.
In the sense of the analog-digital quantum simulation dichotomy~\cite{BN09},
for which a digital quantum simulator can be expressed as a concatenation of a primitive
quantum instruction set comprising, for example, single-qubit gates and a controlled-not (CNOT) gate.
In fact our qubit-channel quantum simulator requires just one CNOT gate and one ancillary qubit,
hence is minimal in
two-qubit gate cost,
whereas ten two-qubit gates and two ancillary qubits are required using standard Stinespring dilation~\cite{MV06}.
Our quantum simulator also uses far fewer single-qubit rotations
than Stinespring dilation would yield~\cite{WBOS13}.

In this article, we report an experimental quantum simulator which is imitated by a decomposition algorithm. Any single qubit channel can be decomposed into a mixture of two quasi-extreme channels. Experimentally, the quasi-extreme channel is implemented by using optical technology and the mixture of two quasi-extreme channels is realized by combining the collected data from two quasi-extreme channels.

The article is organized as follows. Sec. \ref{sec:theory} provides
a theoretical introduction to channel decomposition and numerical simulation of the decomposition algorithm. 
Sec. \ref{sec:expsetup} describes the experimental setup. 
In Sec. \ref{sec:results} we present the experimental results, including four individual quantum noise channels and a simulation of weak measurement process. 
Finally, Sec. \ref{sec:conclusion} contains discussion and conclusions.

\section{theory}
\label{sec:theory}

We construct a quantum channel simulator that transforms a photonic
qubit approximately according to any channel~$\mathcal E$~\cite{WBOS13},
which is a linear, trace-nonincreasing completely positive map that maps
quantum state~$\rho$ to~$\mathcal E(\rho)$.
The approximate experimental channel~${\mathcal E}_\text{exp}$
is guaranteed to map
within a pre-specified error tolerance~$\epsilon$
with respect to $\diamond$-distance
$d_\diamond$,
which is the metric for quantifying
the worst-case distinguishability of the approximate
from the true final state according to trace distance.
Our classical algorithm for designing a qubit-channel simulating circuit
accepts~$\epsilon$
and a $4\times4$ matrix description of~$\mathcal{E}$
as input and yields a
description~$[\mathcal{C}]$ of the photonic circuit~$\mathcal C$ as output
with~$\mathcal C$ comprising single-qubit and a single two-qubit gate plus classical bits.
A single-qubit channel can be expressed as
\begin{equation}
	\mathcal E(\rho)=\sum_{i,j=0}^3\mathcal{E}_{ij}\Xi_i\rho\Xi_j,\;
	\Xi=(\mathds{1},\bm{\sigma})
\end{equation}
with~$\mathds 1$ the $2\times2$ identity operator
and~$\bm{\sigma}=(X,Y,Z)$ the Pauli matrices.

For any single-qubit channel~$\mathcal{E}$,
$p\in [0,1]$ exists such that
\begin{equation}
	\mathcal{E}
		=p \mathcal{E}^\text{e}_1 +(1-p)\mathcal{E}^\text{e}_2
\end{equation}
 for
each $\mathcal{E}^\text{e}_\imath$ a generalized extreme channel~\cite{WBOS13}.
An arbitrary generalized extreme channel $\mathcal{E}^\text{e}$ is specified by two Kraus operators
\begin{equation}
	M_i=R_{\bm{n}}(2\varphi) K_i R_{\bm{m}}(2\delta),\;
	R_{\bm{r}}(2\theta):=\exp(-i\theta \bm{r}\cdot\bm{\sigma})
\label{eq:MKK}
\end{equation}
and
\begin{equation}
\label{eq:extk}
	K_0=
	\begin{pmatrix}\cos\beta &0\\0&\cos\alpha\end{pmatrix}, \quad
	K_1=
	\begin{pmatrix}0&\sin\alpha\\\sin\beta&0
	\end{pmatrix},
\end{equation}
for $0\leq \alpha, \beta \leq 2\pi$.
Furthermore, the Kraus operators $K_i$ can be realized by the circuit shown in Fig.~\ref{fig:setup}a.
In the circuit,
\begin{equation}
	R_y(2\gamma)=\exp(-iY\gamma)=\mathds{1}\cos\gamma-i Y\sin\gamma,
\end{equation}
and
\begin{equation}
	2\gamma_{1,2}=\beta\mp\alpha\pm\frac{\pi}{2}.
\end{equation}

Each~$\mathcal{E}^\text{e}_\imath$ has eight parameters leading to
17 parameters (including~$p$) for arbitrary~$\mathcal E$.
Random~$\mathcal E$ is generated as
a two-qubit partial trace of a three-qubit Haar-random $SU(8)$ matrix.
Decomposing into Kraus operators~(\ref{eq:MKK})
is achieved by guessing
the 17 parameters and then optimizing by reducing the distance between
the trial channel and the desired channel~$\mathcal E$.
When the trial channel~$\mathcal{E}'$ is sufficiently close to~$\mathcal E$,
the optimization routine terminates with the 17-parameter decomposition as output.

We test the decomposing algorithm by numerical simulation. In the numerical simulation, an arbitrary channel is generated from a randomly chosen unitary operator
$U\in SU(8)$ and the channel form can be derived from Kraus operators $K_i=\langle i| U|0\rangle$.
Five examples of input channels are
\begin{widetext}
\begin{align}
  \mathcal{C}_1=&
      \begin{pmatrix}
        0.9276 + 0.0000i  & 0.1125 + 0.0039\text{i}& -0.0027 - 0.0150i & -0.1900 - 0.3383\text{i}\\
   0.1125 - 0.0039\text{i}&  0.4846 + 0.0000i  & 0.0210 + 0.0554i&   0.0787 - 0.1229\text{i}\\
  -0.0027 + 0.0150i  & 0.0210 - 0.0554\text{i}&  0.0724 + 0.0000i & -0.1125 - 0.0039\text{i}\\
  -0.1900 + 0.3383i   &0.0787 + 0.1229i&  -0.1125 + 0.0039i  & 0.5154 + 0.0000i
      \end{pmatrix},\\
  \mathcal{C}_2=&
      \begin{pmatrix}
   0.7187 + 0.0000i&  -0.1056 + 0.1020i &  0.1736 - 0.0679\text{i}& -0.0741 - 0.2666\text{i}\\
  -0.1056 - 0.1020i&   0.8006 + 0.0000i & -0.1944 - 0.1911\text{i}& -0.1028 + 0.1411\text{i}\\
   0.1736 + 0.0679i&  -0.1944 + 0.1911\text{i}&  0.2813 + 0.0000i &  0.1056 - 0.1020\text{i}\\
  -0.0741 + 0.2666i&  -0.1028 - 0.1411\text{i}&  0.1056 + 0.1020i &  0.1994 + 0.0000i
      \end{pmatrix},\\
  \mathcal{C}_3=&
      \begin{pmatrix}
   0.5248 + 0.0000i & -0.2288 + 0.1541\text{i}& -0.2444 - 0.1881\text{i}& -0.2190 + 0.1014i\\
  -0.2288 - 0.1541\text{i}&  0.3227 + 0.0000i &  0.0601 + 0.0484\text{i}&  0.2818 - 0.1173i\\
  -0.2444 + 0.1881\text{i}&  0.0601 - 0.0484\text{i}&  0.4752 + 0.0000i &  0.2288 - 0.1541i\\
  -0.2190 - 0.1014\text{i}&  0.2818 + 0.1173\text{i}&  0.2288 + 0.1541\text{i}&  0.6773 + 0.0000i
      \end{pmatrix},\\
  \mathcal{C}_4=&
      \begin{pmatrix}
   0.3788 + 0.0000i &  0.1571 - 0.1211\text{i}& -0.3581 - 0.0580i &  0.0979 + 0.0813i\\
   0.1571 + 0.1211\text{i}&  0.5037 + 0.0000i & -0.2062 - 0.4069\text{i}&  0.3727 + 0.1591i\\
  -0.3581 + 0.0580i & -0.2062 + 0.4069\text{i}&  0.6212 + 0.0000i & -0.1571 + 0.1211i\\
   0.0979 - 0.0813\text{i}&  0.3727 - 0.1591\text{i}& -0.1571 - 0.1211\text{i}&  0.4963 + 0.0000i
      \end{pmatrix},\\
  \mathcal{C}_5=&
        \begin{pmatrix}
  0.4503 + 0.0000i &  0.1535 - 0.0604\text{i}&  0.2019 + 0.1541\text{i}&  0.0964 + 0.1834\text{i}\\
   0.1535 + 0.0604\text{i}&  0.4329 + 0.0000i & -0.0169 + 0.0529\text{i}& -0.1842 - 0.0254\text{i}\\
   0.2019 - 0.1541\text{i}& -0.0169 - 0.0529\text{i}&  0.5497 + 0.0000i & -0.1535 + 0.0604\text{i}\\
   0.0964 - 0.1834\text{i}& -0.1842 + 0.0254\text{i}& -0.1535 - 0.0604\text{i}&  0.5671 + 0.0000i
      \end{pmatrix}.
\end{align}
\end{widetext}

Our task is to optimize the parameters specifying
the channel decomposition.
Parametrization of unitary operator $U\in SU(2)$ is
\begin{equation}\label{}
  U=\text{e}^{-\text{i} \theta \vec{n}\cdot \vec{\sigma}}=\cos\theta \bm{I} -\text{i}\sin\theta (\vec{n}\cdot \vec{\sigma}),
\end{equation}
For a generalized extreme channel $\mathcal{E}_1^\text{e}$,
the initial rotation $R_{\bm{m}}^{\mathcal{E}_1}(2\delta)$ is parameterized by $\delta^{\mathcal{E}_1}$, $m_{1}^{\mathcal{E}_1}$, $m_{2}^{\mathcal{E}_1}$,
the final rotation $R_{\bm{n}}^{\mathcal{E}_1}(2\varphi)$
by~$\varphi^{\mathcal{E}_1}$, 
$n_1^{\mathcal{E}_1}$, $n_2^{\mathcal{E}_1}$,
and Kraus operators by~$\alpha^{\mathcal{E}_1}$ and $\beta^{\mathcal{E}_1}$.
For generalized extreme channel $\mathcal{E}_2^\text{e}$,
the initial rotation $R_{\bm{m}}^{\mathcal{E}_2}(2\delta)$ is parameterized by $\delta^{\mathcal{E}_2}$, $m_{1}^{\mathcal{E}_2}$, $m_{2}^{\mathcal{E}_2}$,
the final rotation $R_{\bm{n}}^{\mathcal{E}_2}(2\varphi)$
by~$\varphi^{\mathcal{E}_2}$, $n_1^{\mathcal{E}_2}$, $n_2^{\mathcal{E}_2}$
and Kraus operators by $\alpha^{\mathcal{E}_2}$ and $\beta^{\mathcal{E}_2}$.
Simulation results are shown in Table~\ref{tab:sim}.

\begin{table*}
  \centering
\scalebox{1}{
\begin{tabular}{|l|l|l|l|l|l|l|l|l|l|l|l|l|l}
  \hline
           & $\mathcal{C}_1$  & $\mathcal{C}_2$  & $\mathcal{C}_3$  & $\mathcal{C}_4$  & $\mathcal{C}_5$ & & $\mathcal{C}_1$  & $\mathcal{C}_2$  & $\mathcal{C}_3$  & $\mathcal{C}_4$  & $\mathcal{C}_5$ \\ \hline
  $m_1^{\mathcal{E}_1}$      & 0.1043 & 0.5254 & 0.9756 & 0.7444 & 0.2633   & $\beta^{\mathcal{E}_1}$  & 3.3938 & 3.7248 & 3.9264 & 2.5101 & 3.6886 \\ \hline
  $m_2^{\mathcal{E}_1}$      & 0.2658 & 0.7401 & 0.2193 & 0.1878 & 0.2494  & $\alpha^{\mathcal{E}_2}$   & 5.3916 & 2.5514 & 0.7277 & 1.5968 & 1.6137 \\ \hline
  $n_1^{\mathcal{E}_1}$      & 0.3944 & 0.1263 & 0.5393 & 0.9714 & 0.1862  &  $\beta^{\mathcal{E}_2}$    & 3.1851 & 6.2832 & 3.2350 & 4.0102 & 4.0769 \\ \hline
  $n_2^{\mathcal{E}_1}$      & 0.9124 & 0.9920 & 0.4951 & 0.2373 & 0.5166 &  $\delta^{\mathcal{E}_1}$ & 5.6926 & 4.7360 & 2.3722 & 1.7879 & 2.5147 \\ \hline
  $m_1^{\mathcal{E}_2}$      & 0.9768 & 0.0000 & 0.7061 & 0.8913 & 0.0120 &  $\varphi^{\mathcal{E}_1}$ & 6.1559 & 3.1887 & 2.2483 & 0.3786 & 4.5601 \\ \hline
  $m_2^{\mathcal{E}_2}$      & 0.2148 & 0.9502 & 0.6602 & 0.0000 & 0.4274   &  $\delta^{\mathcal{E}_2}$ & 0.1796 & 3.9477 & 0.5693 & 1.4948 & 1.1443 \\ \hline
  $n_1^{\mathcal{E}_2}$      & 0.8214 & 0.2991 & 0.2434 & 0.7347 & 0.4465 &  $\varphi^{\mathcal{E}_2}$ & 3.4528 & 1.7924 & 4.7052 & 4.5230 & 1.7113 \\ \hline
  $n_2^{\mathcal{E}_2}$      & 0.5598 & 0.4585 & 0.7350 & 0.4100 & 0.8948 &  $p$        & 0.4675 & 0.7001 & 0.8370 & 0.9032 & 0.4922 \\ \hline
  $\alpha^{\mathcal{E}_1}$ & 3.8118 & 4.7982 & 6.1211 & 5.8303 & 5.4589 &  $\epsilon$ & 0.0031 & 0.0009 & 0.0040 & 0.0006 & 0.0045 \\
  \hline
\end{tabular}}
\caption{%
	Numerical simulation of the channel decomposition for five randomly chosen input channels
and channel~$\mathcal{C}_T$.
	Error $\epsilon$ is the actual error from the simulation.%
}
\label{tab:sim}
\end{table*}

\section{experimental setup}
\label{sec:expsetup}
\begin{figure}[h!]
\centering
\includegraphics[width=\linewidth]{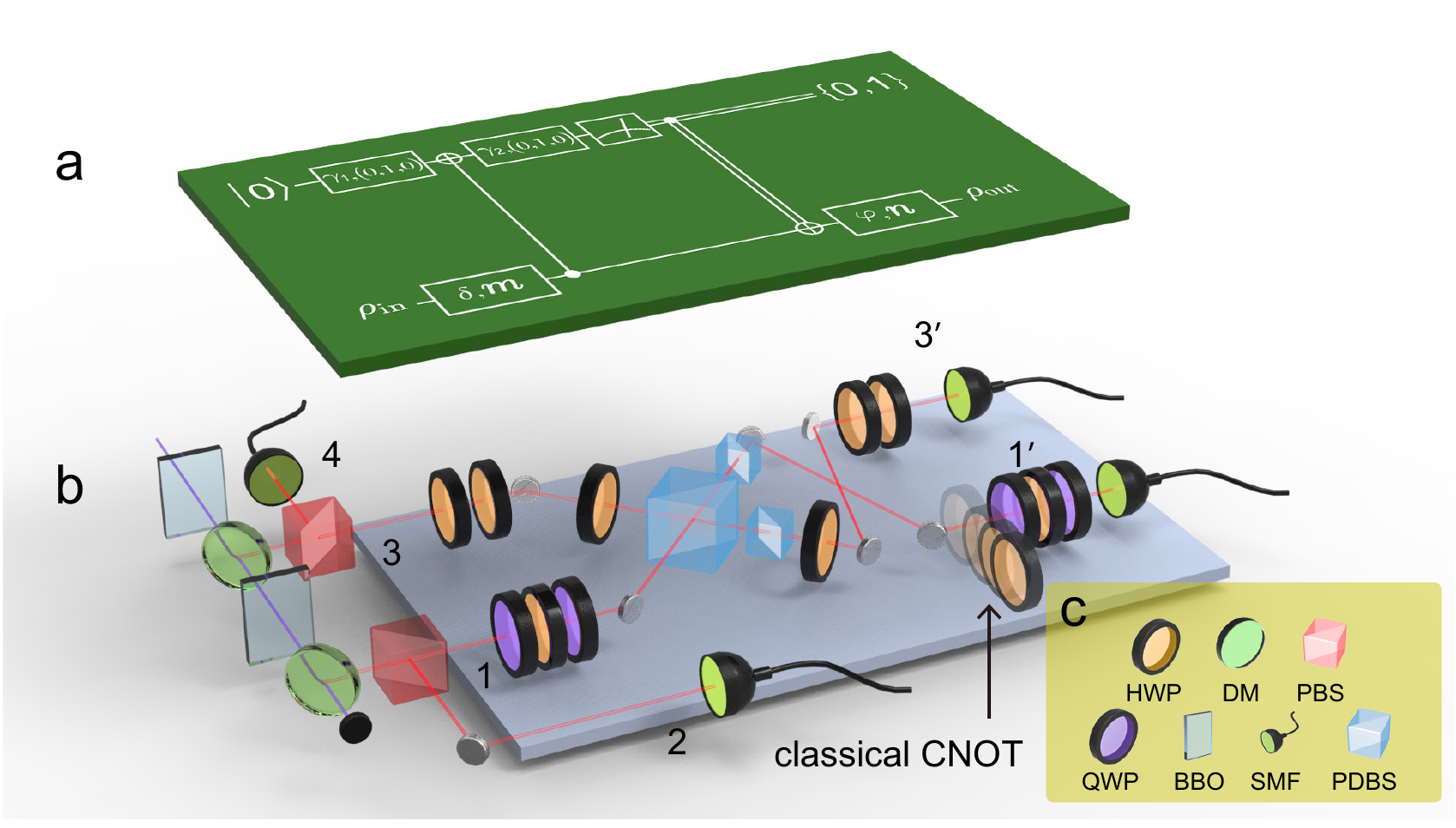}
\caption{\textbf{Experimental scheme for quantum channel simulation. a, }
The quantum circuit for~$M_i$ (\ref{eq:MKK}),
comprising single-qubit rotations, one quantum CNOT gate, and one classical CNOT gate.
When the measurement result on ancilla is~$\ket{0}$ ($\ket{1}$), the circuit
postselectively acts as $M_0$ ($M_1$).
(b)~Schematic drawing of the experimental setup.
(c)~
Symbols used in \textbf{b}.  }
\label{fig:setup}
\end{figure}

Realization of Kraus operators $M_i$~(\ref{eq:MKK}) is described by the circuit shown in Fig.~\ref{fig:setup}a
and implemented according to the schematic of Fig.~\ref{fig:setup}b. A femtosecond pulse (150fs, 80MHz, 780nm) is converted to ultraviolet pulses (390nm) through a frequency doubler $\text{LiB}_3\text{O}_5$ crystal.
Then the ultraviolet pulse (150fs, 80MHz, 390 nm) passes through two 2-mm-thick collinear BBO crystals,
creating two pairs of photons~$\ket{HV}_{ij}$ with central wavelength of 780nm. 
The ultraviolet pulse (390nm) and the generated photons (780nm) are along the same direction and separated by a dichroic mirror (DM).

The generated photons~$\ket{HV}_{ij}$ are separated by a PBS and the reflected photons are detected to guarantee that the transmitted photons are underway.
All four photons are collected by the SMF and detected by the SPCM.
All the photons are filtered by narrowband interference filters with $\Delta \lambda_\text{FWHM}= 2.8$nm (fullwidth at half-maximum) prior to detection.
Throughout the entire experiment,
the two-fold coincidence rates
for $\ket{HV}_{12}$ and $\ket{HV}_{34}$ are $9\times 10^{4} \text{s}^{-1}$ and $10\times 10^{4} \text{s}^{-1}$, respectively.
Overall detective efficiency is approximately~19$\%$.
We use a homemade Field Programmable Gate Array (FPGA) to record the fourfold coincidence (not shown here).

The single-photon rotation gates are realized by the combination of half-wave plates (HWPs) and quarter-wave plates (QWPs). 
The effect of HWP and QWP whose fast axes are at angles $\tau$ and $\xi$ with respect to the vertical axis, respectively, are given by the 2$\times$2 matrices,
\begin{equation}
\begin{split}
U_\text{HWP}(\tau)=\begin{pmatrix}
	\cos2\tau & -\sin2\tau \\
	-\sin2\tau & -\cos2\tau
\end{pmatrix}, \\
U_\text{QWP}(\xi)=\begin{pmatrix}
1+\text{i}\cos2\xi & -\text{i}\sin2\xi\\
 -\text{i}\sin2\xi & 1-\text{i}\cos2\xi
\end{pmatrix}/\sqrt{2}
\label{eq:WPmatrix}
\end{split}
\end{equation}

The rotation around the~$y$ axis by angle $2\gamma_{i}$ is
\begin{equation}
	R_{\bm{y}}(2\gamma_{i})
		=\text{e}^{-\text{i}\gamma_{i}\bm{Y}},
		=\cos\gamma_{i} \mathds{1}-\text{i}\sin\gamma_{i}\bm{Y}.
\end{equation}
The combinational operation of two HWPs set at 0$^{\circ}$ and $\tau$, respectively, is in the form
\begin{equation}
U_\text{HWP}(0^{\circ})\cdot U_\text{HWP}(\tau)=\begin{pmatrix}
\cos2\tau & -\sin2\tau \\
\sin2\tau & \cos2\tau
\end{pmatrix}.
\end{equation}
We set $\tau=\gamma_{i}/2$;
then the operation is $R_{\bm{y}}(2\gamma_{i})$.

The rotation around
\begin{equation}
	\bm{r}\cdot\bm{\sigma}=AX+BY+CZ
\end{equation}
by angle $2\theta$, where $A, B$ and $C$ satisfy $A^2+B^2+C^2=1$, can be implemented by a HWP set at angle $\tau$ sandwiched by two QWPs set at angles $\xi_{1}$ and $\xi_{2}$, respectively. The combinational operation of the three wave plates is 
\begin{equation}
\begin{split}
U_\text{QWP}(\xi_{1})U_\text{HWP}(\tau)U_\text{QWP}(\xi_{2})\\
=\begin{pmatrix}
\cos\theta-iC\sin\theta & -\sin\theta(iA+B)\\
-\sin\theta(iA-B)            & \cos\theta-iC\sin\theta
\end{pmatrix}
\end{split}
\end{equation}
where
\begin{align}
	\cos\theta
		=&\cos\Theta\cos\Lambda,\;
	A
		=\sin\Theta\cos\Lambda/\sqrt{\Omega},
		\nonumber\\
	B
		=&\cos\Theta\sin\Lambda/\sqrt{\Omega},\;
	C
		=\sin\Theta\sin\Lambda/\sqrt{\Omega}
\end{align}
with
\begin{equation}
	\Theta=\xi_{1}-\xi_{2},\;\Lambda=2\tau-\xi_{1}-\xi_{2},\;
	\Omega=1-\cos^2\Theta\cos^2\Lambda.
\end{equation}
By appropriately choosing the angles $\tau$, $\xi_{1}$ and $\xi_{2}$, rotation $R_{\bm{r}}(2\theta)$  can be implemented. 

The two-photon controlled NOT (CNOT) gate~\cite{BPW03} are realized by overlapping two photons on a polarization-dependent beamsplitter (PDBS).
The system and ancilla photonic qubits are generated by shining the ultraviolet pluses
on two collinear $\beta$-barium borate (BBO) crystals emitting photon pairs~$\ket{HV}_{ij}$ along the pumping direction,
with~$\ket{H}$ and~$\ket{V}$ the horizontal- and vertical-polarization states,
and~$i$,~$j$ denote the path mode.
The generated photons in the pair~$\ket{HV}_{ij}$  are separated by a polarizing beam splitter (PBS),
which transmits the~$\ket{H}$ component and reflects the~$\ket{V}$ component for each photon.
Reflected photons~2 and~4 are collected by single-mode fibers (SMFs)
and detected by single-photon counting modules (SPCMs)
to herald that photons~1 (system) and~3 (ancilla) are underway, respectively~\cite{KMW95}.

\begin{figure*}
\includegraphics[width=1.8\columnwidth]{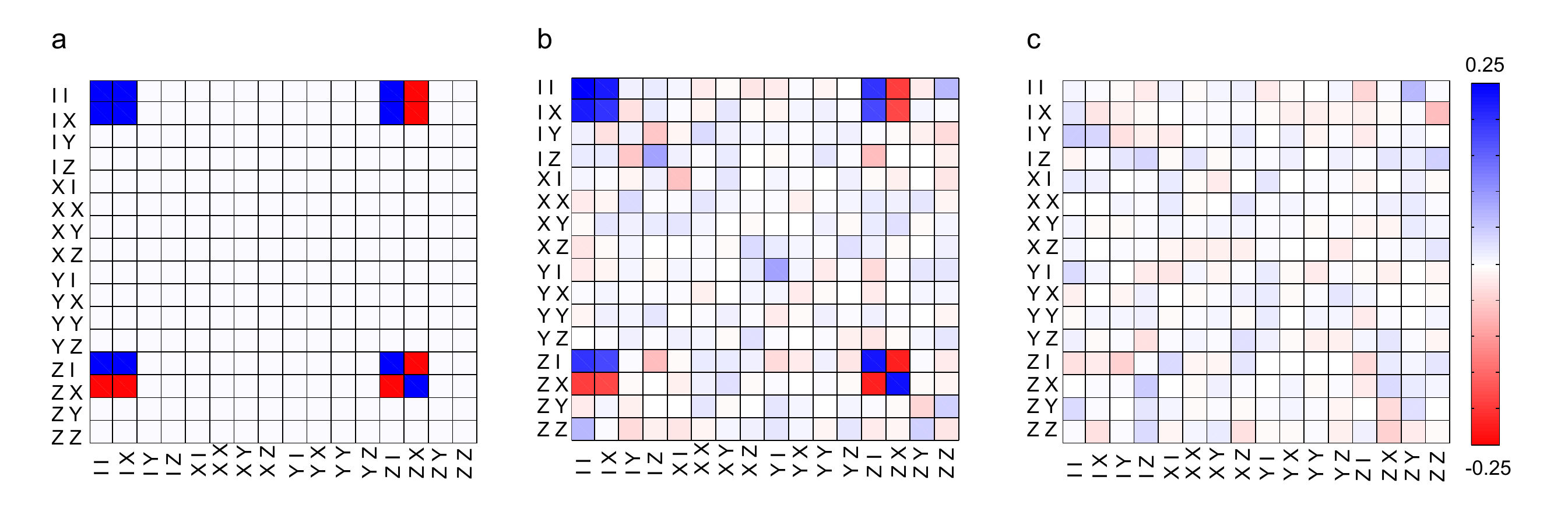}
\caption{\textbf{Process matrix of the CNOT operation. a.} Real elements of ideal process matrix
(imaginary elements are identically zero).
\textbf{b. } The real elements of~$\chi_\text{exp}$. \textbf{c. } Imaginary elements of~$\chi_\text{exp}$. }
\label{CNOTmatrix}
\end{figure*}

We experimentally characterize the quantum CNOT gate
via quantum process tomography (QPT) technology~\cite{OPG04} and obtain
gate fidelity $F=0.83\pm0.02$ as shown in Fig.\ref{CNOTmatrix}

The classical CNOT operation is a classical logic operation
that flips the system-qubit state~$1^{\prime}$
conditioned on the measurement result of ancillary qubit $3^{\prime}$.
Experimentally, classical CNOT is effectively statistically simulated:
we set the measurement basis of ancillary photon $3^{\prime}$ on~$\ket{H}$ or~$\ket{V}$
with equal probability.
No further operation on system qubit~$1^{\prime}$ occurs
when the measurement basis choice of ancillary photon $3^{\prime}$ is~$\ket{H}$,
whereas an $X$ operation (an HWP set at $45^{\circ}$) is applied on the system qubit $1^{\prime}$
when the measurement basis choice of ancillary photon $3^{\prime}$ is~$\ket{V}$.
If the ancilla-qubit measurement result is~$\ket{H}$ ($\ket{V}$), the simulator is described by~$M_0$ ($M_1$).

The probability~$p$ is also statistically simulated.
We first set up the circuit for simulating~$\mathcal E^\text{e}_1$
and collect data for time~$t_1$.
Then we convert the circuit to the case of simulating~$\mathcal E^\text{e}_2$
and collect data for time~$t_2$.
Combining these data yields
$\mathcal{E}(\rho)
	=(t_1 \mathcal E^\text{e}_1 +t_2 \mathcal E^\text{e}_2)/(t_1+t_2)$.
By choosing~$t_1$ and~$t_2$ appropriately,
any $p\in[0,1]$ can thus be simulated.

We emphasise here that the rotations in our experiment are realized manually,
and the classical CNOT gate is implemented by inserting an HWP according to the projector choice on photon $3^{\prime}$. The mixture of $\mathcal E^\text{e}_1$ and $\mathcal E^\text{e}_2$ is statistically simulated by collecting data from $\mathcal E^\text{e}_1$ and $\mathcal E^\text{e}_2$ with different times. In our experiment, simulating one single-qubit channel needs us to run the setup four times and then combine the collected data. 

\begin{figure}[h!]
\includegraphics[width=1\columnwidth]{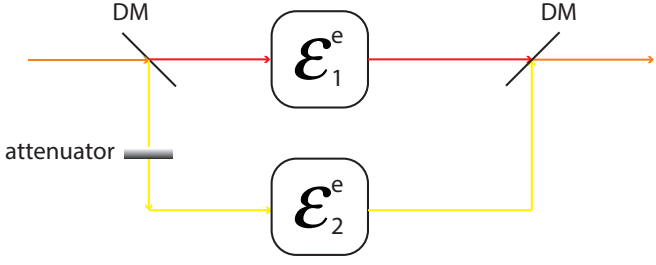}
\caption{\textbf{Proposed experimental scheme to simulate single-qubit channel with one run. }}
\label{fig:proposal}
\end{figure}

In fact, the four runs can be embedded into one run. Here, we also propose an experimental scheme to simulate any single-qubit channel within one run. Below we summarize the experimental scheme. As shown in Fig.~\ref{fig:proposal},
a beam of light with two different colors, with central wavelength of $\lambda_1$ and $\lambda_2$ respectively, 
is split by a dichroic mirror(DM). The transmitted color $\lambda_1$ go through  $\mathcal E^\text{e}_1$ as shown in Fig.~\ref{fig:setup}. 
The reflected color $\lambda_1$ first go to an automatically controlled attenuator and then go through $\mathcal E^\text{e}_1$. 
The attenuator setting depends on the parameter $p\in[0,1]$. 
For example, the attenuator block half of $\lambda_2$ is equivalence of simulating $p=2/3$.
Finally, $\lambda_1$ and $\lambda_2$ is recombined on another DM and injected to the detector.
All the rotations can be realized by an automatically controlled waveplate and the classical CNOT can be replaced by feedforward technology.

\section{Results}
\label{sec:results}
\subsection{Random channel }
We first show that a randomly chosen channel is accurately simulated with our setup.
Randomly chosen input channel $\mathcal{E}$
\begin{widetext}
\begin{equation}
	\mathcal{E}=\begin{pmatrix}
			0.3938  &	0.0075+0.0739\text{i}& 0.0172+0.0155\text{i}& 0.0801-0.0614i \\
			0.0075-0.0739\text{i}& 0.1594 & -0.0733-0.0801\text{i}& -0.066+0.0172i \\
			0.0172-0.0155\text{i}& -0.0733+0.0801\text{i}& 0.2241 & -0.014-0.075i \\
			0.0801+0.0614\text{i}& -0.066-0.0172\text{i}& -0.014+0.075\text{i}& 0.2228
	\end{pmatrix}
\label{eq:arbitrarymatrix}
\end{equation}
\end{widetext}
is realized by the circuit in Fig.~\ref{fig:setup}b with appropriate
parameters specified. 
The decomposition of $\mathcal{E}$ is shown in Table~\ref{tab:random}.
For $p=0.6$, we set~$t_1=60 \text{s}$ and~$t_2=40 \text{s}$.

\begin{table}
  \centering
\begin{tabular}{|p{0.4\linewidth}|p{0.45\linewidth}|}
  \hline
  \hline
  $m_1^{\mathcal{E}_1}$      & 0.1896  \\ \hline
  $m_2^{\mathcal{E}_1}$      & 0.7948  \\ \hline
  $n_1^{\mathcal{E}_1}$      & -0.7813   \\ \hline
  $n_2^{\mathcal{E}_1}$      & 0.5804  \\ \hline
  $m_1^{\mathcal{E}_2}$      & 0.3901  \\ \hline
  $m_2^{\mathcal{E}_2}$      & 0.5051   \\ \hline
  $n_1^{\mathcal{E}_2}$      & -0.0919 \\ \hline
  $n_2^{\mathcal{E}_2}$      & 0.9817  \\ \hline
  $\alpha^{\mathcal{E}_1}$ & 0.18$\pi$  \\ \hline
  $\beta^{\mathcal{E}_1}$  & 0.26$\pi$   \\ \hline
  $\alpha^{\mathcal{E}_2}$   & 0.84$\pi$  \\ \hline
  $\beta^{\mathcal{E}_2}$    & 0.40$\pi$  \\ \hline
  $\delta^{\mathcal{E}_1}$ & 0.42$\pi$  \\ \hline
  $\varphi^{\mathcal{E}_1}$ & 0.36$\pi$  \\ \hline
  $\delta^{\mathcal{E}_2}$ & -0.75$\pi$  \\ \hline
  $\varphi^{\mathcal{E}_2}$ & 0.56$\pi$ \\ \hline
  $p$        &0.6 \\ \hline
    \hline
\end{tabular}
\caption{Decomposition of channel $\mathcal{E}$
}
\label{tab:random}
\end{table}

\begin{figure}[h!]
\centering
\includegraphics[width=\linewidth]{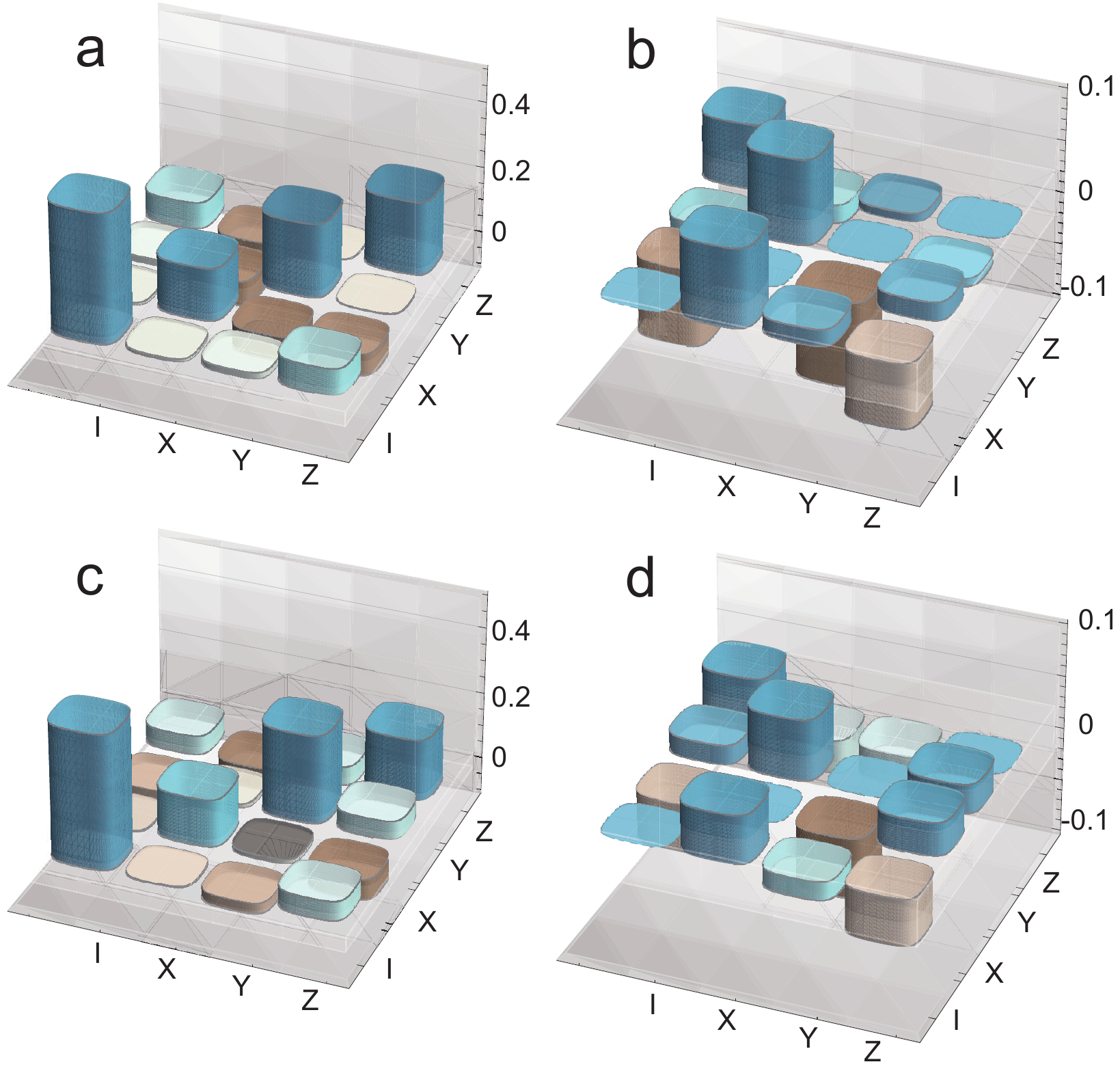}
\caption{%
	Reconstructed process matrix of the randomly chosen channel in Eq.~(\ref{eq:arbitrarymatrix}).
	(a)~The real part of the ideal channel $\mathcal{E}$.
	(b)~The imaginary part of the ideal channel $\mathcal{E}$.
	(c)~The real part of the experimentally constructed channel $\mathcal{E}_\text{exp}$.
	(d)~The imaginary part of the experimentally constructed channel $\mathcal{E}_\text{exp}$. }
\label{fig:arbitrarychannel}
\end{figure}

To verify accurate channel simulation,
we use QPT to reconstruct the matrix representation of~$\mathcal E$.
Figure~\ref{fig:arbitrarychannel} shows the experimentally reconstructed
$\mathcal{E}_\text{exp}$ matrix.
We calculate the process fidelity 
\begin{equation}
	F_{\text{P}}=\operatorname{Tr}\left(\sqrt{\sqrt{\mathcal{E}}\mathcal{E}_\text{exp}\sqrt{\mathcal{E}}}\right)^2
\end{equation}
between the reconstructed matrix
$\mathcal{E}_\text{exp}$ and~$\mathcal{E}$
and discover $F_{\text{P}}=0.94\pm0.02$.

Average fidelity is~\cite{BOS+02}
\begin{equation}
\bar{F}=(2F_{\text{P}}+1)/3=0.96\pm0.01.
\end{equation}
As further analysis, we calculate the trace distance
\begin{equation}
	D(\mathcal{E},\mathcal{E}_\text{exp})
		=\operatorname{Tr}|\mathcal{E}-\mathcal{E}_\text{exp}|/2=0.22\pm0.02.
\end{equation}
Fidelity $F_{\text{P}}$ is related to $D$ by the inequality~\cite{NC00}
\begin{equation}
	1-\sqrt{F_{\text{P}}}\leq D \leq \sqrt{1-F_{\text{P}}}.
\end{equation}
In our case ($F_{\text{P}}=0.94$), the upper and lower bounds of $D$ are 0.06 and 0.24.
\begin{figure*}
\begin{minipage}{0.2\linewidth}
  \centerline{\includegraphics[width=1.2\linewidth]{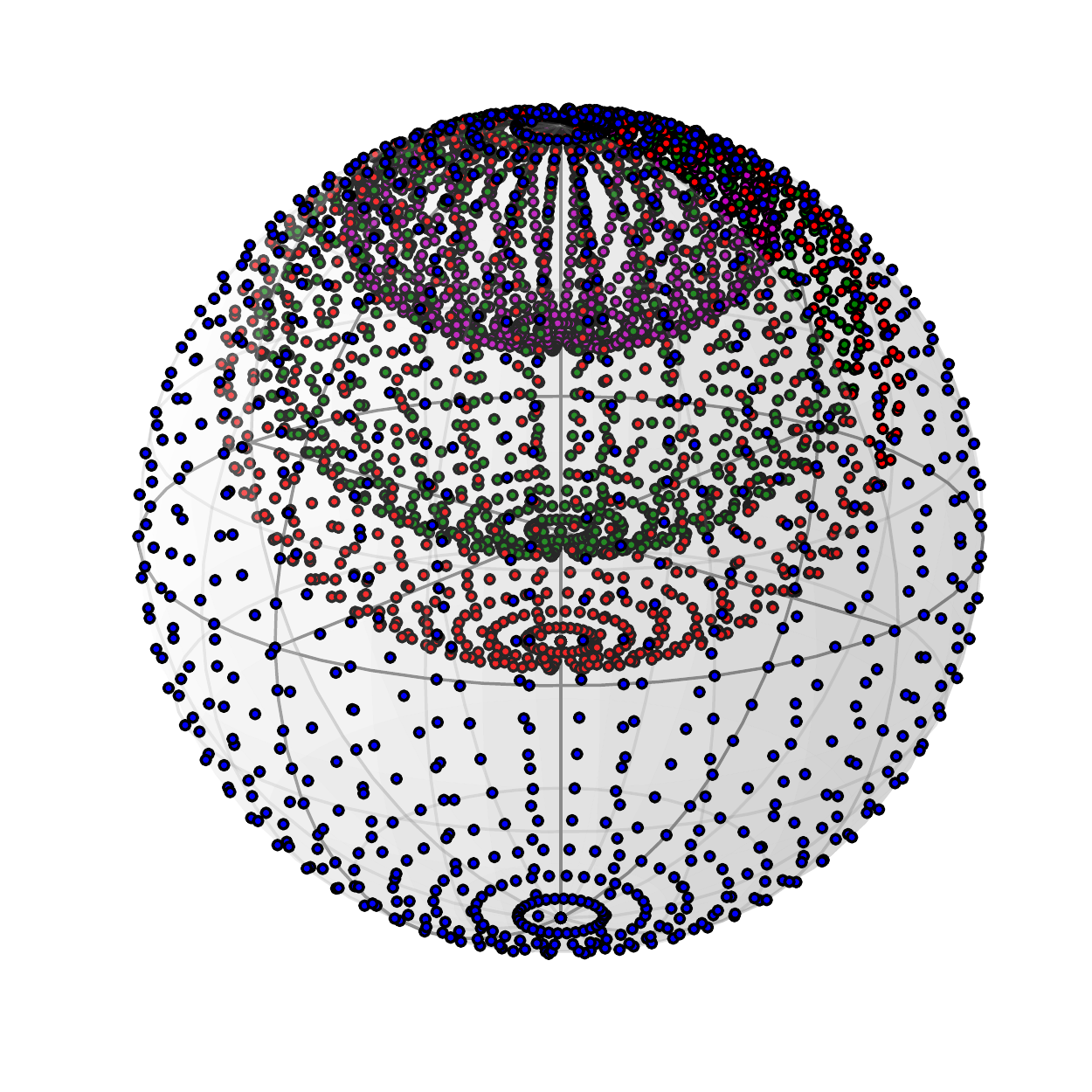}}
  \centerline{(a)}
\end{minipage}
\hfill
\begin{minipage}{0.2\linewidth}
  \centerline{\includegraphics[width=1.2\linewidth]{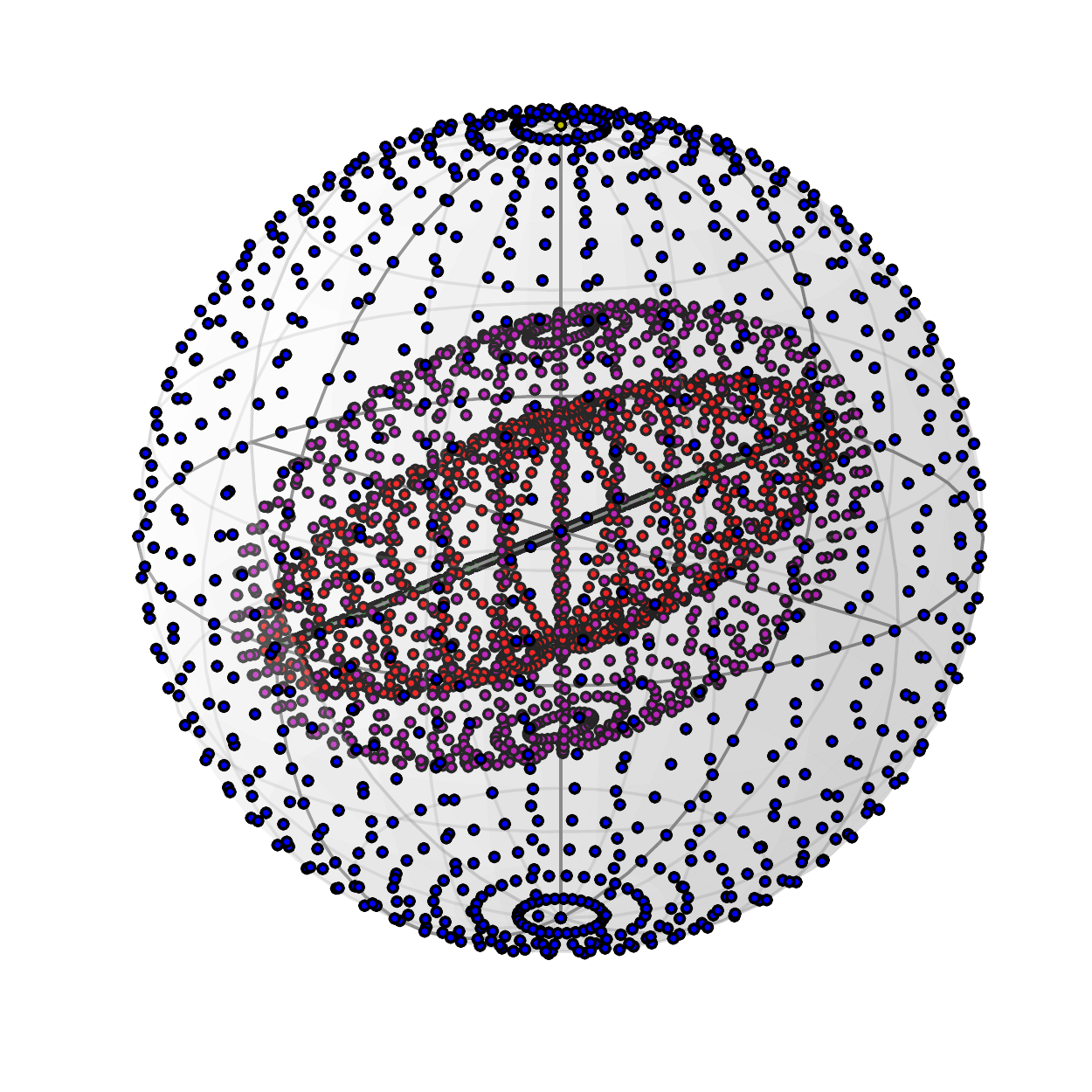}}
  \centerline{(b)}
\end{minipage}
\hfill
\begin{minipage}{0.2\linewidth}
  \centerline{\includegraphics[width=1.2\linewidth]{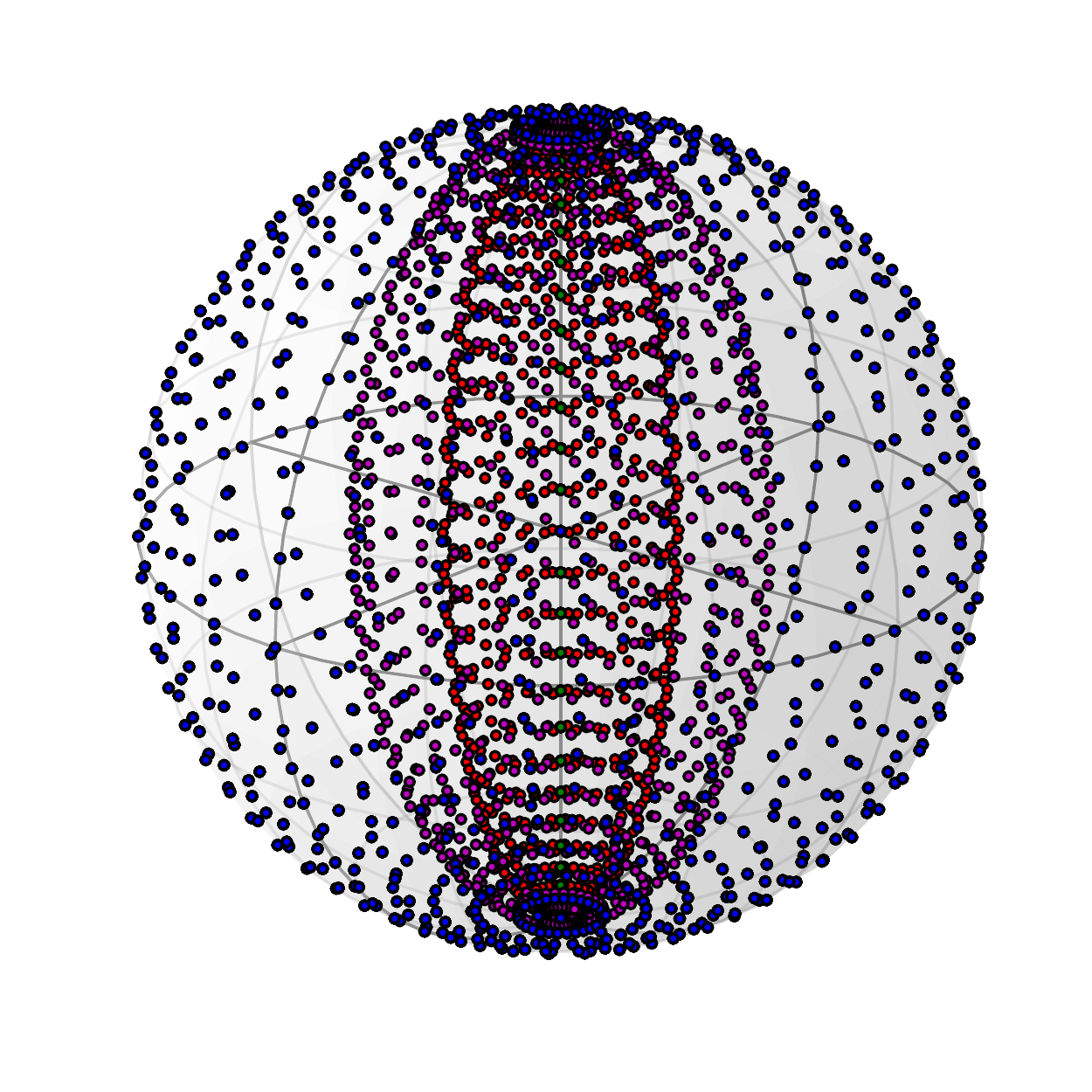}}
  \centerline{(c)}
\end{minipage}
\hfill
\begin{minipage}{0.2\linewidth}
  \centerline{\includegraphics[width=1.2\linewidth]{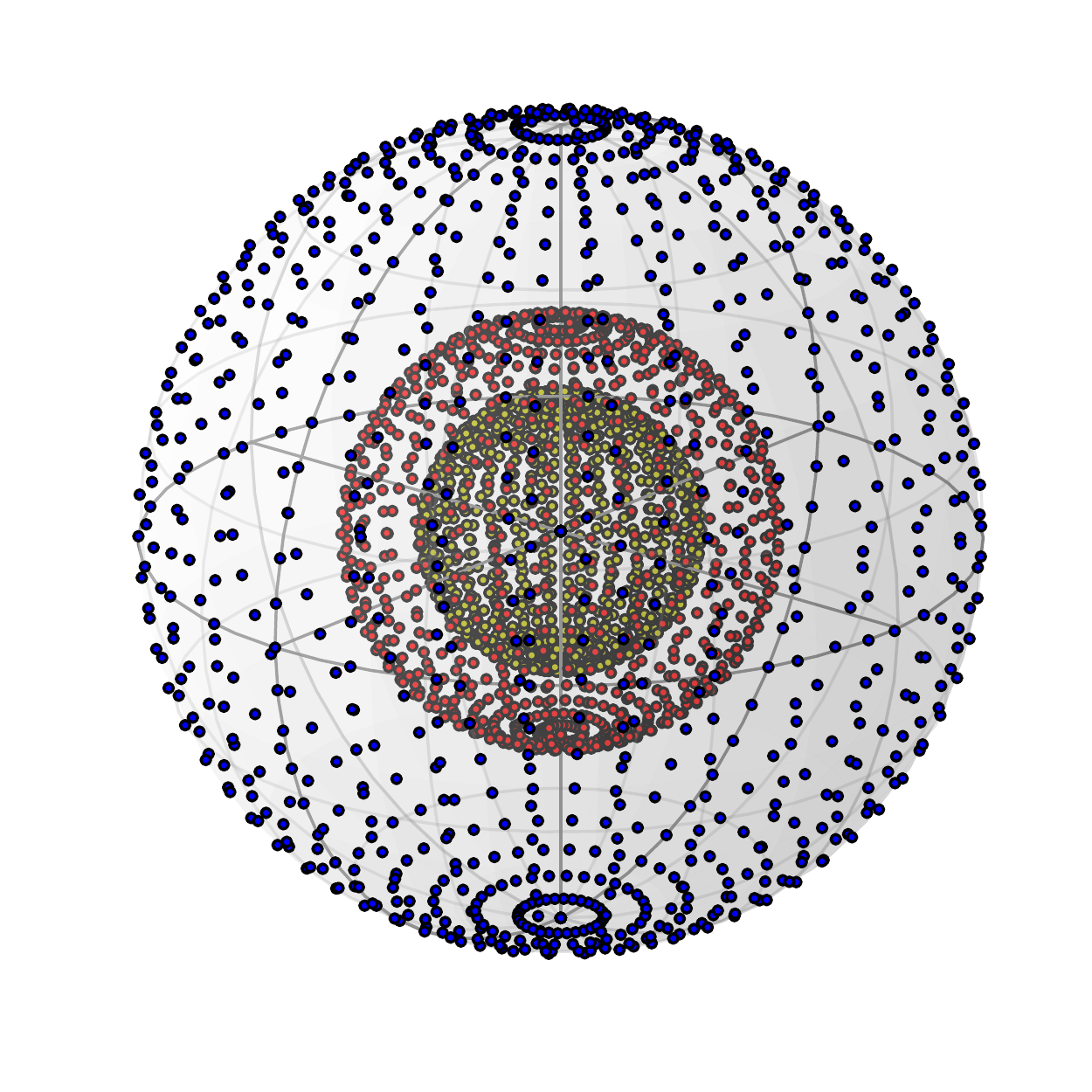}}
  \centerline{(d)}
  \end{minipage}
\hfill
\begin{minipage}{0.2\linewidth}
  \centerline{\includegraphics[width=1.2\linewidth]{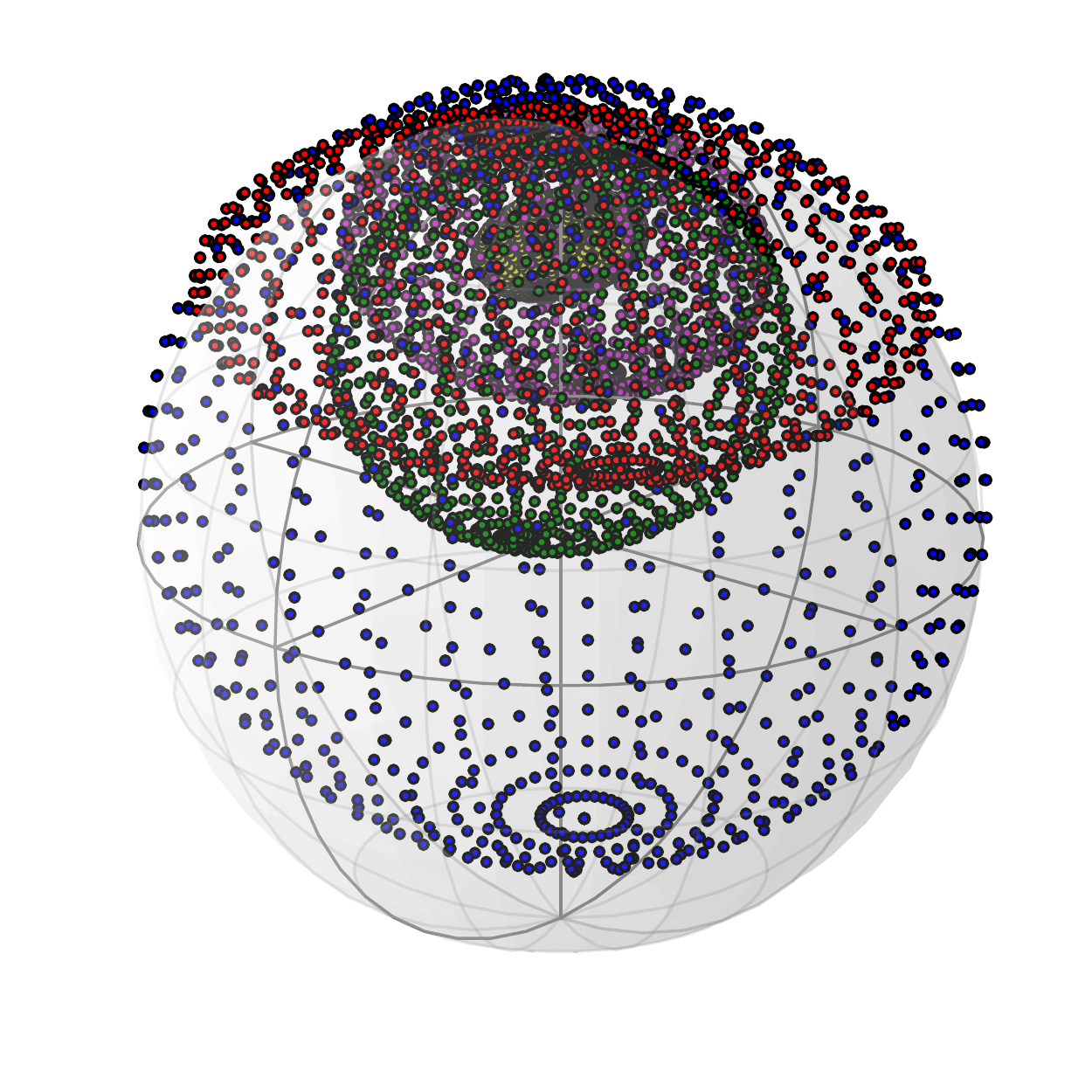}}
  \centerline{(e)}
  \end{minipage}
\hfill
\begin{minipage}{0.2\linewidth}
  \centerline{\includegraphics[width=1.2\linewidth]{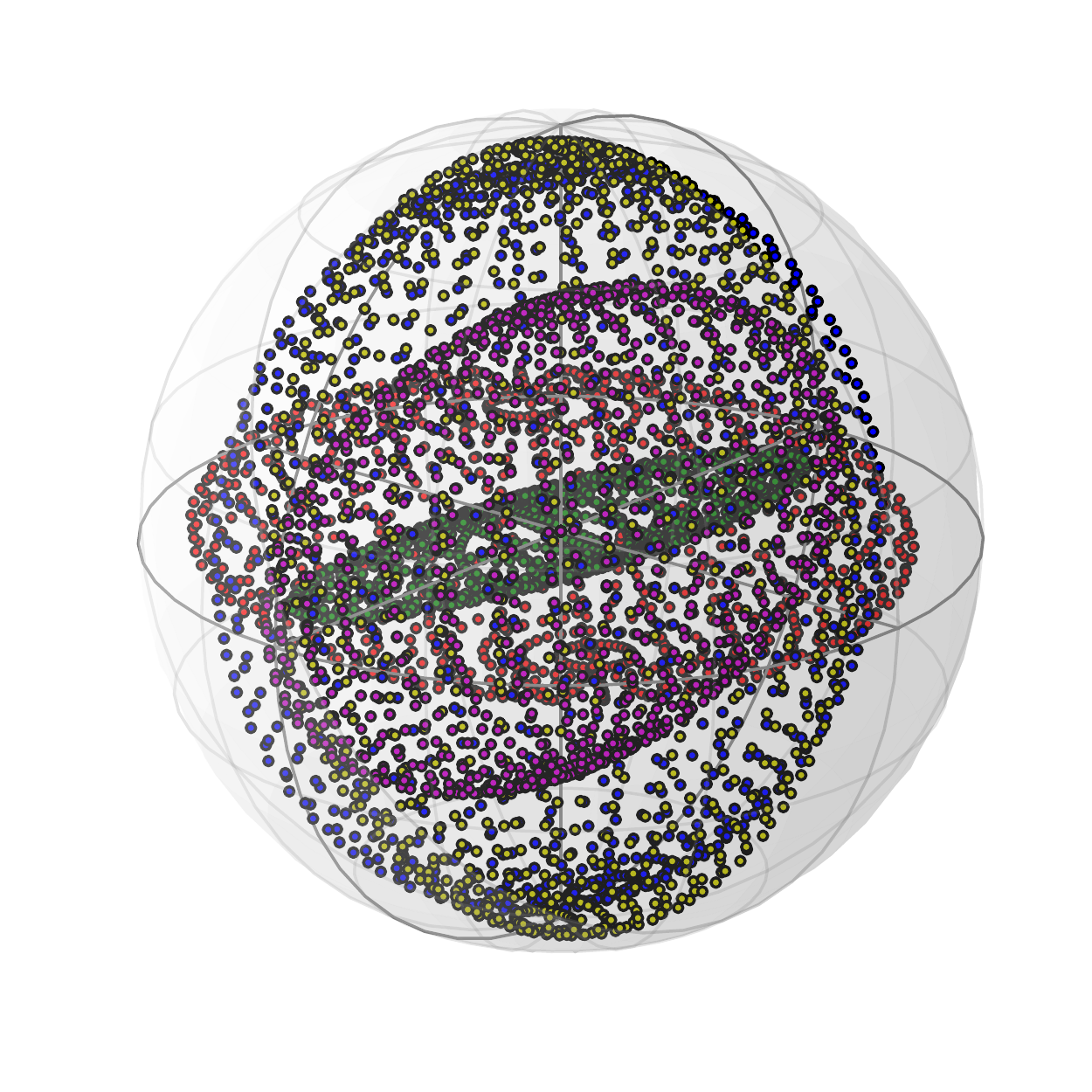}}
  \centerline{(f)}
  \end{minipage}
\hfill
\begin{minipage}{0.2\linewidth}
  \centerline{\includegraphics[width=1.2\linewidth]{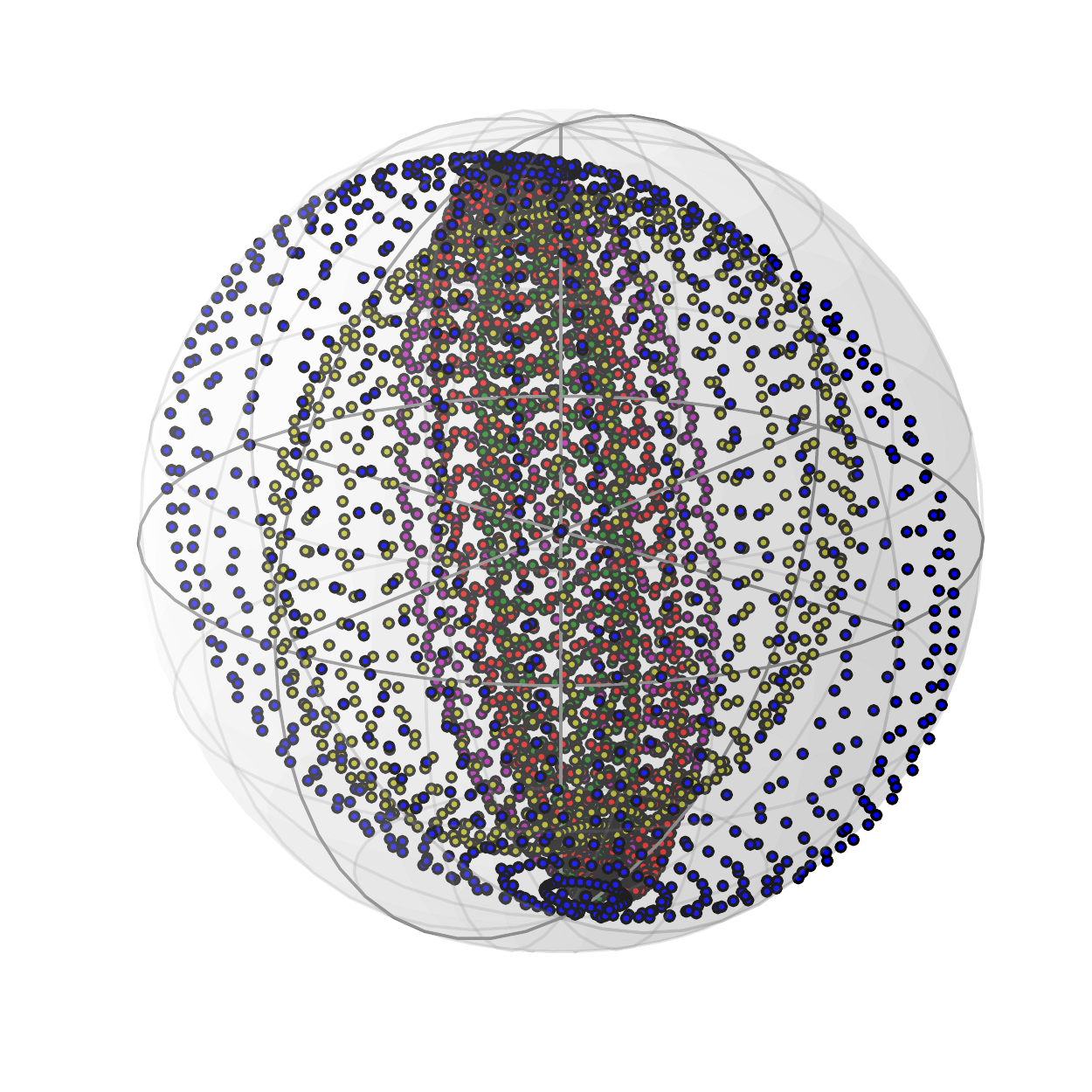}}
  \centerline{(g)}
  \end{minipage}
\hfill
\begin{minipage}{0.2\linewidth}
  \centerline{\includegraphics[width=1.2\linewidth]{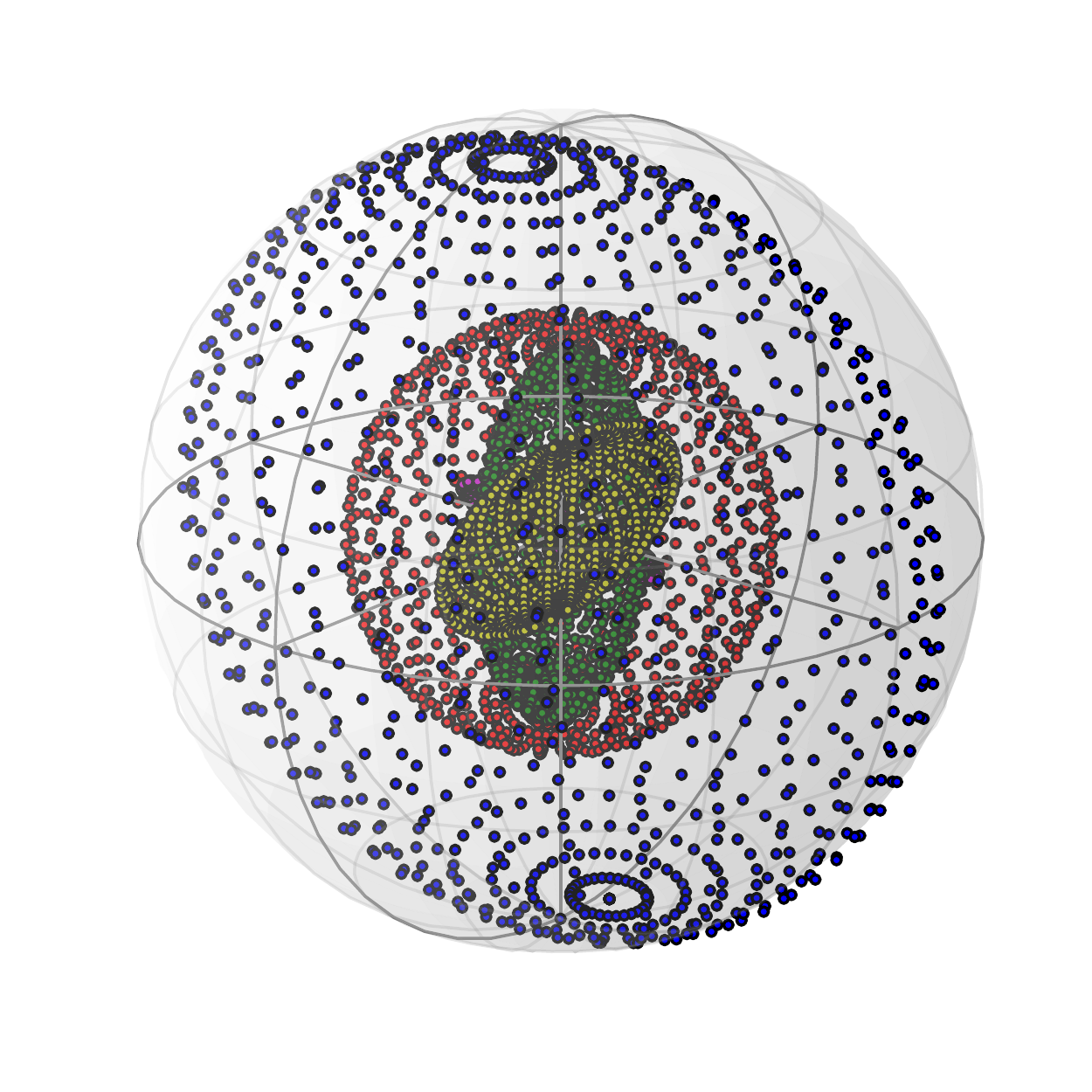}}
  \centerline{(h) }        
\end{minipage}
\caption{
	The geometric interpretation of the quantum-noise channels on a Bloch sphere. 
	(a)--(d), the ideal geometric interpretations of amplitude damping channel, bit-flip channel, phase-flip channel and depolarizing channel with $\lambda=0$(blue dots), $\lambda=0.36$(red dots), $\lambda=0.5$(green dots), $\lambda=0.75$(purple dots) and $\lambda=1$(yellow dots). (e)--(f), the geometric interpretations of experimentally reconstructed channels.
	 }
\label{bloch}
\end{figure*}

\subsection{Amplitude Damping Channel}

The amplitude damping (AD),
or decay channel can be determined by two Kraus operators
\begin{equation}
K_0=\begin{pmatrix}
	1 & 0 \\
	0 & \sqrt{1-\lambda}
\end{pmatrix},
K_1=\begin{pmatrix}
0 & \sqrt{\lambda} \\
0 & 0
\end{pmatrix}.
\label{eq:amdchannekraus}
\end{equation}

Table~\ref{ampparametertable} shows the setting of parameters of the simulator.
\begin{table}[h!]
\begin{center}
\begin{tabular}{|p{0.11\linewidth}|p{0.11\linewidth}|p{0.11\linewidth}|p{0.11\linewidth}|p{0.11\linewidth}|p{0.11\linewidth}|p{0.11\linewidth}|p{0.11\linewidth}|}
\hline

\hline
\multirow{2}{*}{$\lambda$} & \multicolumn{6}{c|}{$\mathcal{E}_1^{\text{e}}$} &\multirow{2}{*}{p} \\\cline{2-7}
 & $\alpha$ & $\beta$ & $\gamma_1$ & $\gamma_{2}$  & $R_{\textbf{m}}(2\delta)$ & $R_{\textbf{n}}(2\varphi)$ & \\\hline
0         & 0            & 0          &  $\pi$/4         & -$\pi$/4 & none & none  & 1        \\\hline
0.36      & 0.103$\pi$   & 0          &  0.2$\pi$        &  -0.2$\pi$  & none & none  & 1     \\\hline
0.5       & $\pi$/4      & 0          & 3$\pi$/8         & -3$\pi$/8     & none & none   & 1     \\\hline
0.75      & $\pi$/3      & 0          & $\pi$/12         & -$\pi$/12     & none & none   & 1    \\\hline
1         & $\pi$/2      & 0          & 0                & 0            & none & none   & 1     \\\hline

\end{tabular}
\end{center}
\caption{Parameters of amplitude damping channel for the given~$\lambda$.}
\label{ampparametertable}
\end{table}
Fig.~\ref{bloch}a(e) shows the geometric interpretation of ideal(experimental) AD channel for $\lambda\in\{0, 0.36, 0.5, 0.75,1\}$.

\subsection{Bit-flip channel}
The bit-flip channel has two Kraus operators in the form
\begin{equation}
\begin{split}
&K_0=\sqrt{1-\lambda}\mathds{1}=
\begin{pmatrix}
\sqrt{1-\lambda} & 0 \\
0 & \sqrt{1-\lambda}
\end{pmatrix},\\
&K_1=\sqrt{\lambda}X=
\begin{pmatrix}
	0 & \sqrt{\lambda} \\
\sqrt{\lambda} & 0
\end{pmatrix}.
\end{split}
\end{equation}
For different~$\lambda$, the corresponding channel parameters are shown in Table~\ref{bitflipparametertable}.
Fig.~\ref{bloch}b(f) shows the geometric interpretation of ideal(experimental) bit-flip channel for $\lambda\in\{0, 0.36, 0.5, 0.75,1\}$.

\begin{table}[h!]
\begin{center}
\begin{tabular}{|p{0.11\linewidth}|p{0.11\linewidth}|p{0.11\linewidth}|p{0.11\linewidth}|p{0.12\linewidth}|p{0.11\linewidth}|p{0.11\linewidth}|p{0.1\linewidth}|}
\hline

\hline
\multirow{2}{*}{$\lambda$} & \multicolumn{6}{c|}{$\mathcal{E}_1^{\text{e}}$} &\multirow{2}{*}{p} \\\cline{2-7}
 & $\alpha$ & $\beta$ & $\gamma_1$  & $\gamma_{2}$ &$R_{\textbf{m}}(2\delta)$ & $R_{\textbf{n}}(2\varphi)$ & \\\hline
0         & 0            & 0           &  $\pi$/4          & -$\pi$/4   & none & none  & 1        \\\hline
0.36      & 0.103$\pi$   & 0.103$\pi$  &  $\pi$/4          &  -0.044$\pi$     & none & none  & 1  \\\hline
0.5       & $\pi$/4      & $\pi$/4     &  $\pi$/4          & 0       & none & none  & 1           \\\hline
0.75      & $\pi$/3      & $\pi$/3     &  $\pi$/4          & $\pi$/12     & none & none  & 1      \\\hline
1         & $\pi$/2      & $\pi$/2     &  $\pi$/4          & $\pi$/4    & none & none  & 1        \\\hline

\end{tabular}
\end{center}
\caption{Parameters of bit-flip channel for the given~$\lambda$.}
\label{bitflipparametertable}
\end{table}

\subsection{Phase-flip channel}
The phase-flip channel has two Kraus operators in the form,
\begin{equation}
\begin{split}
&K_0=\sqrt{1-\lambda}\mathds{1}=
\begin{pmatrix}
\sqrt{1-\lambda} & 0 \\
0 & \sqrt{1-\lambda}
\end{pmatrix},\\
&K_1=\sqrt{\lambda}Z=
\begin{pmatrix}
\sqrt{\lambda} & 0 \\
0 & -\sqrt{\lambda}
\end{pmatrix}.
\end{split}
\label{eq:phaseflipKraus}
\end{equation}

For different~$\lambda$, the setting of the parameters are shown in Table~\ref{phaseflipparametertable}. Fig.~\ref{bloch}c(g) shows the geometric interpretation of ideal(experimental) phase-flip channel for $\lambda\in\{0, 0.36, 0.5, 0.75,1\}$.

\begin{table*}
\begin{tabular}{|p{0.05\linewidth}|p{0.05\linewidth}|p{0.05\linewidth}|p{0.05\linewidth}|p{0.05\linewidth}|p{0.05\linewidth}|p{0.05\linewidth}|p{0.05\linewidth}|p{0.05\linewidth}|p{0.05\linewidth}|p{0.05\linewidth}|p{0.05\linewidth}|p{0.05\linewidth}|p{0.05\linewidth}|}
\hline

\hline
\multirow{2}{*}{$\lambda$} & \multicolumn{6}{|c|}{$\mathcal{E}_1^{\text{e}}$} &  \multicolumn{6}{|c|}{$\mathcal{E}_{2}^{\text{e}}$} & $p$ \\\cline{2-13}
&$\alpha$ & $\beta$ & $\gamma_1$  & $\gamma_{2}$ & $R_{\textbf{m}}(\delta)$  &$R_{\textbf{n}}(\psi)$ & $\alpha$ &$\beta$ & $\gamma_1$  & $\gamma_{2}$ & $R_{\textbf{m}}(\delta)$  &$R_{\textbf{n}}(\psi)$ &  \\\hline
0         & 0            & 0           &  $\pi$/4          & -$\pi$/4     & none & none    &    $\pi$     &   0         & -$\pi$/4          &   $\pi$/4     & none & none   & 1   \\\hline
0.36      & 0            & 0           &  $\pi$/4          & -$\pi$/4    & none & none     &    $\pi$     &   0         & -$\pi$/4          &   $\pi$/4   & none & none     & 0.64\\\hline
0.5       & 0            & 0           &  $\pi$/4          & -$\pi$/4     & none & none    &    $\pi$     &   0         & -$\pi$/4          &   $\pi$/4     & none & none   & 0.5 \\\hline
0.75      & 0            & 0           &  $\pi$/4          & -$\pi$/4    & none & none     &    $\pi$     &   0         & -$\pi$/4          &   $\pi$/4    & none & none    & 0.25 \\\hline
1         & 0            & 0           &  $\pi$/4          & -$\pi$/4     & none & none    &    $\pi$     &   0         & -$\pi$/4          &   $\pi$/4    & none & none    &  1   \\\hline

\end{tabular}
\caption{Parameters of phase-flip channel for the given~$\lambda$.}
\label{phaseflipparametertable}
\end{table*}

\subsection{Depolarizing channel}
The depolarizing channel, which is known as a white-noise channel,
has the form
\begin{equation}
\mathcal{E}(\rho)=(1-\lambda)\rho+\frac{\lambda}{3}(X\rho X+Y\rho Y+Z\rho Z).
\end{equation}
The setting of the parameters are shown in Table~\ref{depolarizationparametertable}.
Fig.~\ref{bloch}e(h) shows the geometric interpretation of ideal(experimental) depolarizing channel for $\lambda\in\{0, 0.36, 0.5, 0.75,1\}$.

\begin{table*}
\begin{tabular}{|p{0.05\linewidth}|p{0.05\linewidth}|p{0.05\linewidth}|p{0.05\linewidth}|p{0.05\linewidth}|p{0.05\linewidth}|p{0.05\linewidth}|p{0.05\linewidth}|p{0.05\linewidth}|p{0.05\linewidth}|p{0.05\linewidth}|p{0.05\linewidth}|p{0.05\linewidth}|p{0.05\linewidth}|}
\hline

\hline
\multirow{2}{*}{$\lambda$} & \multicolumn{6}{|c|}{$\mathcal{E}_1^{\text{e}}$} &  \multicolumn{6}{|c|}{$\mathcal{E}_{2}^{\text{e}}$} & $p$ \\\cline{2-13}
&$\alpha$ & $\beta$ & $\gamma_1$  & $\gamma_{2}$ & $R_{\textbf{m}}(\delta)$  &$R_{\textbf{n}}(\psi)$ & $\alpha$ &$\beta$ & $\gamma_1$  & $\gamma_{2}$ & $R_{\textbf{m}}(\delta)$  &$R_{\textbf{n}}(\psi)$ &  \\\hline
0         & 0            & 0           &  $\pi$/4          & -$\pi$/4    & none & none     &    $\pi$/4   &   $\pi$/4   & $\pi$/4           &   0        & Y  & none      & 1   \\\hline
0.36      & 0.13$\pi$    & 0.13$\pi$   &  $\pi$/4          & -0.12$\pi$/   & none & none   &    $\pi$/4   &   $\pi$/4   & $\pi$/4           &   0    & Y & none          & 0.76\\\hline
0.5       & $\pi$/6      & $\pi$/6     &  $\pi$/4          & -$\pi$/12    & none & none    &    $\pi$/4   &   $\pi$/4   & $\pi$/4           &   0       & Y & none       & 0.66 \\\hline
0.75      & $\pi$/4      & $\pi$/4     &  $\pi$/4          & 0        & none & none        &    $\pi$/4   &   $\pi$/4   & $\pi$/4           &   0       & Y & none       & 0.5 \\\hline
1         & $\pi$/2      & $\pi$/2     &  $\pi$/4          & $\pi$/4    & none & none      &    $\pi$/4   &   $\pi$/4   & $\pi$/4           &   0      & Y & none        & 0.33   \\\hline

\end{tabular}
\caption{Parameters of depolarizing channel for the given~$\lambda$.}
\label{depolarizationparametertable}
\end{table*}

\subsection{Weak measurement}

\begin{figure*}[h!]
\centering
\includegraphics[width=\linewidth]{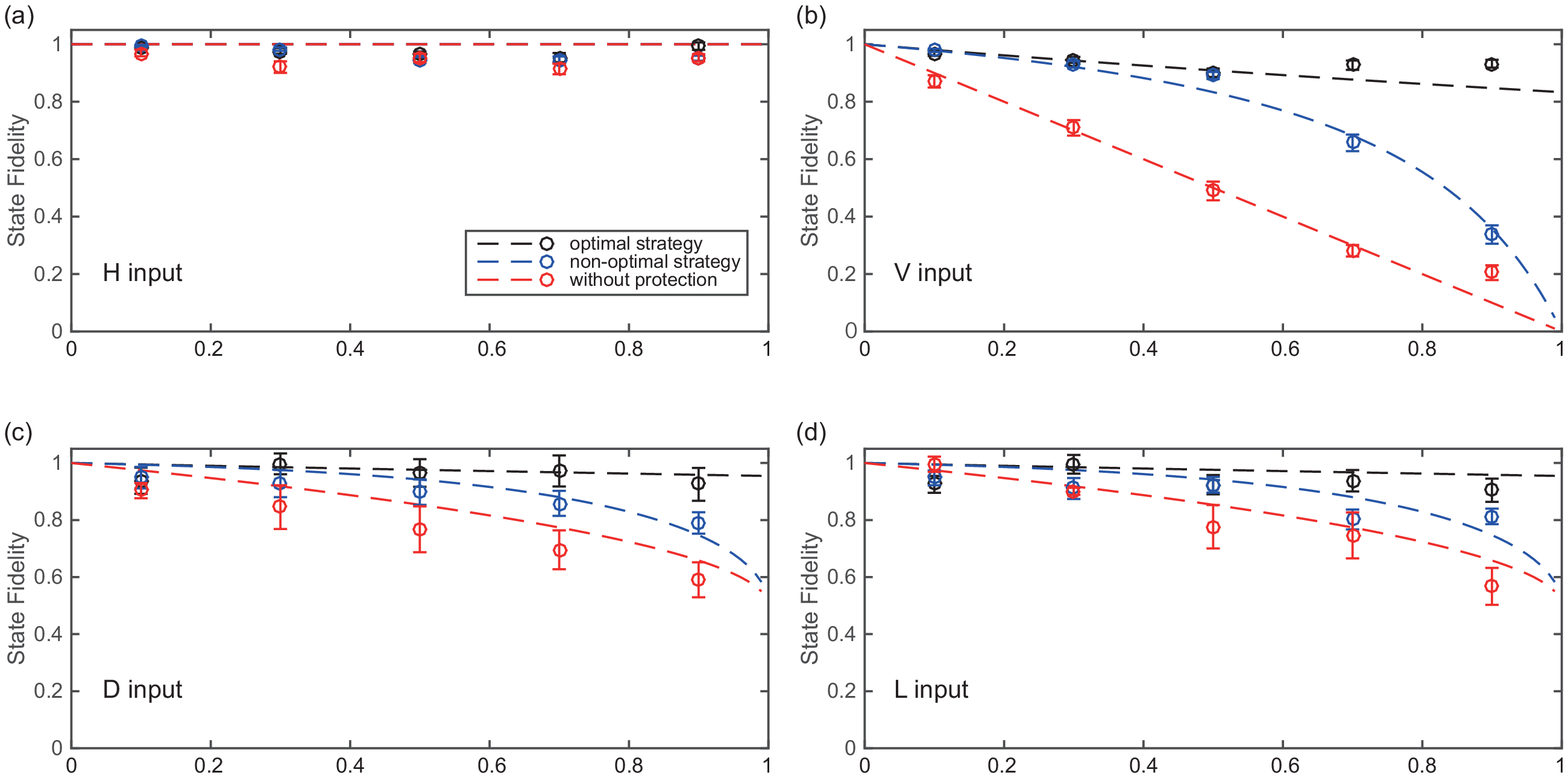}
\caption{%
	State-fidelity dynamics for the amplitude-damping channel with weak measurement protection.
	Dashed lines represent theoretical results, and dots represent experimental results.
	Red, blue and black colors represent no weak measurement,
	non-optimal measurement strategy ($p_1=p_2$)
	and optimal measurement strategy ($p_2=p_1+\lambda(1-p_1)$), respectively.
	The input state is
	(a)~$\ket{H}$,
	(b)~$\ket{V}$,
	(c)~$\ket{D}$, and
	(d)~$\ket{L}$
  }
\label{weakmeasurestate}
\end{figure*}

We show that our apparatus successfully simulates trace-decreasing channels,
such as weak measurement followed by measurement reversal,
which is a strategy for offsetting amplitude damping $\mathcal{E}_\text{AD}$~(\ref{eq:amdchannekraus})
at the cost of losing particles through postselection~\cite{Korotkov2010,Kim2011, Kim09}.
For a single-qubit input state~$\rho$,
this strategy is
\begin{equation}
	\rho\mapsto\rho'=N[\mathcal{E}_\text{AD}(M\rho M^{\dag})]N^{\dag}
\end{equation}
for weak measurement
$M=\operatorname{diag}(1,\sqrt{1-p_1})$ with $p_1\sim0$
and weak measurement reversal
$N=\operatorname{diag}(\sqrt{1-p_2},1)$ with $p_2\sim0$.
A successful outcome corresponds to high fidelity
$\operatorname{Tr}\left(\sqrt{\sqrt{\mathcal{\rho'}}\mathcal{\rho}\sqrt{\mathcal{\rho'}}}\right)^2$
with success probability $\operatorname{Tr}\rho'$.

Larger~$p_1$ corresponds to superior protection
and smaller success probability.
Seeking to explore the trade-off between success probability and weak measurement strength~$p_1$,
we choose a fairly strong measurement strength $p_1=0.8$
and then let $p_2=p_1+\lambda(1-p_1)$ if damping parameter~$\lambda$ is given
(this relation between~$p_1$ and~$p_2$ is the ``optimal strategy'');
otherwise $p_2=p_1$ if~$\lambda$ is unknown
(the ``non-optimal strategy'')~\cite{wangshuchao2014}.
Theoretical and experimental state fidelity results
for input states
$\ket{H}$,~$\ket{V}$,~$\ket{D}=1/\sqrt{2}(\ket{H}+\ket{V})$
and~$\ket{L}=1/\sqrt{2}(\ket{H}+\text{i}\ket{V})$
are shown in Fig.~\ref{weakmeasurestate}
for three cases: pure amplitude damping,
non-optimal measurement strategy,
and optimal measurement strategy.
Note that~$\ket{H}$ input is immune to $\mathcal{E}_\text{AD}$,
but, due to experimental imperfection, the fidelity for input state $\ket{H}$ is not exactly 1.
We find that the optimal strategy provides the best protection,
and the experimental results agree
with the theory for all three cases. 

\section{conclusion}
\label{sec:conclusion}
In this article, we demonstrate that a digital channel simulator can be realized via linear optics. 
Any open-system quantum dynamics and quantum channels on single qubit can be simulated in our system.
For multi-qubit channel simulation, decomposition algorithm has been extended to qudit channels\cite{WS15}. 
In large-scale channel simulation, linear optics system might be retarded by the probabilistic CNOT gate. 
However, other systems, such as superconducting qubit and trapped ions, can benefit from our results.
Our demonstration can serve as a foundation for future experimental simulations employing
networks of qubit channel simulators.
Such networks could serve to simulate general dissipative many-body dynamics
including the interplay between dissipative and unitary processes~\cite{DDG+10}
and dissipative universal quantum computation~\cite{VWC09}
by combining two-qubit entangling gates with the qubit-channel simulators.

\begin{acknowledgments}
We acknowledge insightful discussions with I. Dhand, W.-J. Zou, Y. Chen and H.-H. Wang.
This work has been supported by the National Natural Science Foundation of China,
the Chinese Academy of Sciences,
and the National Fundamental Research Program (grant no.\ 11404318 and no.\ 2011CB921300). 
H. L. was partially supported by Shanghai Sailing Program.
X.-C.\ Y.\ was also supported by the Alexander von Humboldt Foundation.
and B.C.S.\ acknowledges financial support from the 1000 Talent Plan, NSERC and AITF.
\end{acknowledgments}

\bibliography{Reference}
\end{document}